\documentclass{desyproc}

\usepackage{graphicx}

\newcommand{\bgam}{\beta_\gamma}
\newcommand{\bmat}{\beta_\mathrm{m}}

\newcommand{\DVloop}{\Delta V_\mathrm{1-loop}}
\newcommand{\eotwash}{E\"ot-Wash}
\newcommand{\meff}{m_\mathrm{eff}}       
\newcommand{\Mpl}{M_\mathrm{Pl}} 
\newcommand{\phibulk}{\phi_\mathrm{B}}   
\newcommand{\Veff}{V_\mathrm{eff}} 

\begin{document}

\title{Particles and forces from chameleon dark energy}
\author{\slshape{Amol Upadhye}\\[1ex]
Argonne National Laboratory, 9700 S. Cass Ave., Lemont, IL 60439}%

\maketitle

\begin{abstract}
Chameleon dark energy is a matter-coupled scalar field which hides its fifth forces locally by becoming massive.  We estimate torsion pendulum constraints on the residual fifth forces due to models with gravitation-strength couplings.  Experiments such as E\"ot-Wash are on the verge of ruling out ``quantum-stable'' chameleon models, in which quantum corrections to the chameleon field and mass remain small.  We also consider photon-coupled chameleons, which can be tested by afterglow experiments such as CHASE. 
\end{abstract}

\section{Introduction}
\label{sec:introduction}

The accelerating expansion of the universe is well-supported by the data~\cite{Komatsu_etal_2010,Suzuki_etal_2012,Sanchez_etal_2012}, but its cause is the greatest mystery in modern cosmology.  Dynamical alternatives to a ``cosmological constant'' density $\rho_\Lambda \approx 10^{-120}\Mpl^4$ explain the smallness of $\rho_\Lambda$ by fields tunneling among local minima of their potential~\cite{Bousso_Polchinski_2000,Steinhardt_Turok_2006}, or by a slow decrease of the vacuum energy known as ``degravitation''~\cite{Dvali_Hofmann_Khoury_2007,deRham_Hofmann_Khoury_Tolley_2008}.  
At low energies, the simplest of these models reduce to effective ``dark energy'' scalar fields which may evolve with time or couple to known particles.  Matter-coupled scalars mediate fifth forces which must be screened at high densities in order to evade local constraints~\cite{Adelberger_etal_2009}.  The three best-understood screened dark energy models are: chameleons, which acquire large effective masses~\cite{Khoury_Weltman_2004a,Khoury_Weltman_2004b,Brax_etal_2004}; symmetrons, which decouple from matter through a symmetry-restoring phase transition~\cite{Hinterbichler_Khoury_2010,Olive_Pospelov_2007}; and Galileons, in which a non-canonical kinetic energy term reduces the effective matter coupling~\cite{Nicolis_Rattazzi_Trincherini_2008}.  
We will discuss ``quantum-stable'' chameleon models, in which the one-loop Coleman-Weinberg corrections to the field and mass remain small~\cite{Upadhye_Hu_Khoury_2012}.

On the experimental front, the dark energy scales $M_\Lambda = \rho_\Lambda^{1/4} \sim 10^{-3}$~eV and $1/M_\Lambda \sim 100$~$\mu$m are readily accessible in the laboratory~\cite{Adelberger_Heckel_Nelson_2003}. 
We demonstrate that the \eotwash~torsion pendulum experiment~\cite{Kapner_etal_2007} is on the verge of excluding quantum-stable chameleons with gravitation-strength matter couplings.  A coupling between dark energy and electromagnetism would imply that dark energy particles could be produced through photon oscillation in a magnetic field.  
We show that such particles can be trapped by afterglow experiments including CHASE~\cite{Steffen_etal_2010,Upadhye_Steffen_Weltman_2010,Upadhye_Steffen_Chou_2012}.
This paper is organized as follows.  Section~\ref{sec:fifth_forces} discusses chameleon fifth forces and quantum stability.  Oscillation and afterglow constraints are covered in Sec.~\ref{sec:new_particles}.

\section{Fifth forces}
\label{sec:fifth_forces}

We begin with a scalar field coupled to the trace of the matter stress tensor $-T_\mu^\mu \approx \rho$, and possibly to the electromagnetic field strength tensor $F_{\mu\nu}$, with effective potential
$
\Veff(\phi)
=
V(\phi) 
-
\bmat T_\mu^\mu \phi / \Mpl
+ \bgam F_{\mu\nu}F^{\mu\nu}\phi / (4\Mpl).
$
The self-interaction $V(\phi)$ can be approximated as a constant plus a power law when specific examples are necessary~\cite{Brax_etal_2004}.  In this section we assume a static system with $\bgam=0$, reducing the equation of motion $\Box\phi = \Veff'(\phi)$ to $\nabla^2\phi = V'(\phi) + \bmat\rho/\Mpl$. Inside a constant-density bulk the spatial derivatives vanish, so $\phi$ takes its bulk value $\phibulk$ defined by $V'(\phibulk) = -\bmat\rho/\Mpl$.  With $-V',\,V'',\,-V'''>0$, an increase in $\rho$ implies a  lower $\phibulk$, hence a larger effective mass $\meff(\phi) \equiv \Veff''(\phi)^{1/2}$ (the ``chameleon effect''). 

\begin{figure}[tb]
  \includegraphics[angle=270,width=2.9in]{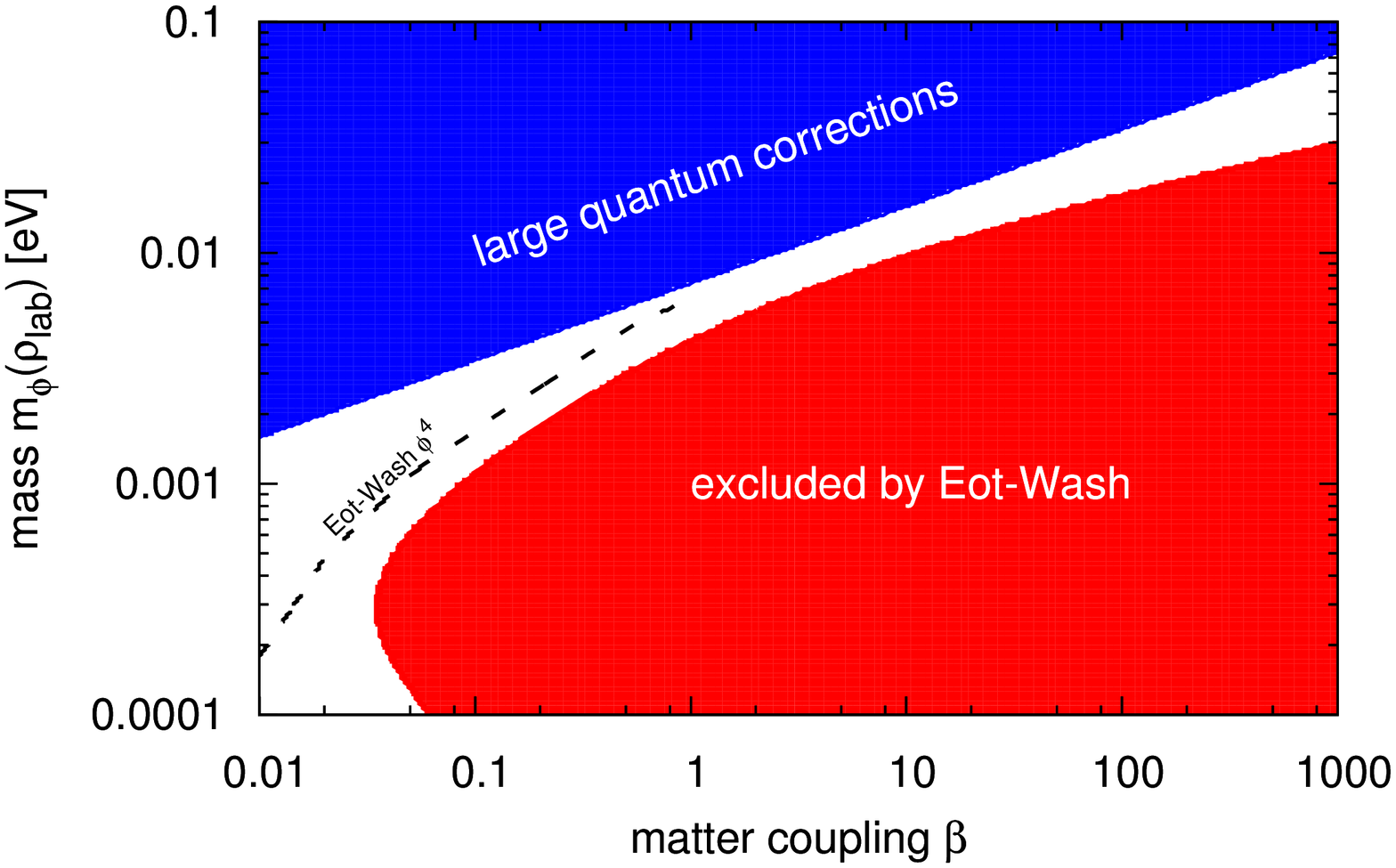}
  \includegraphics[angle=270,width=2.7in]{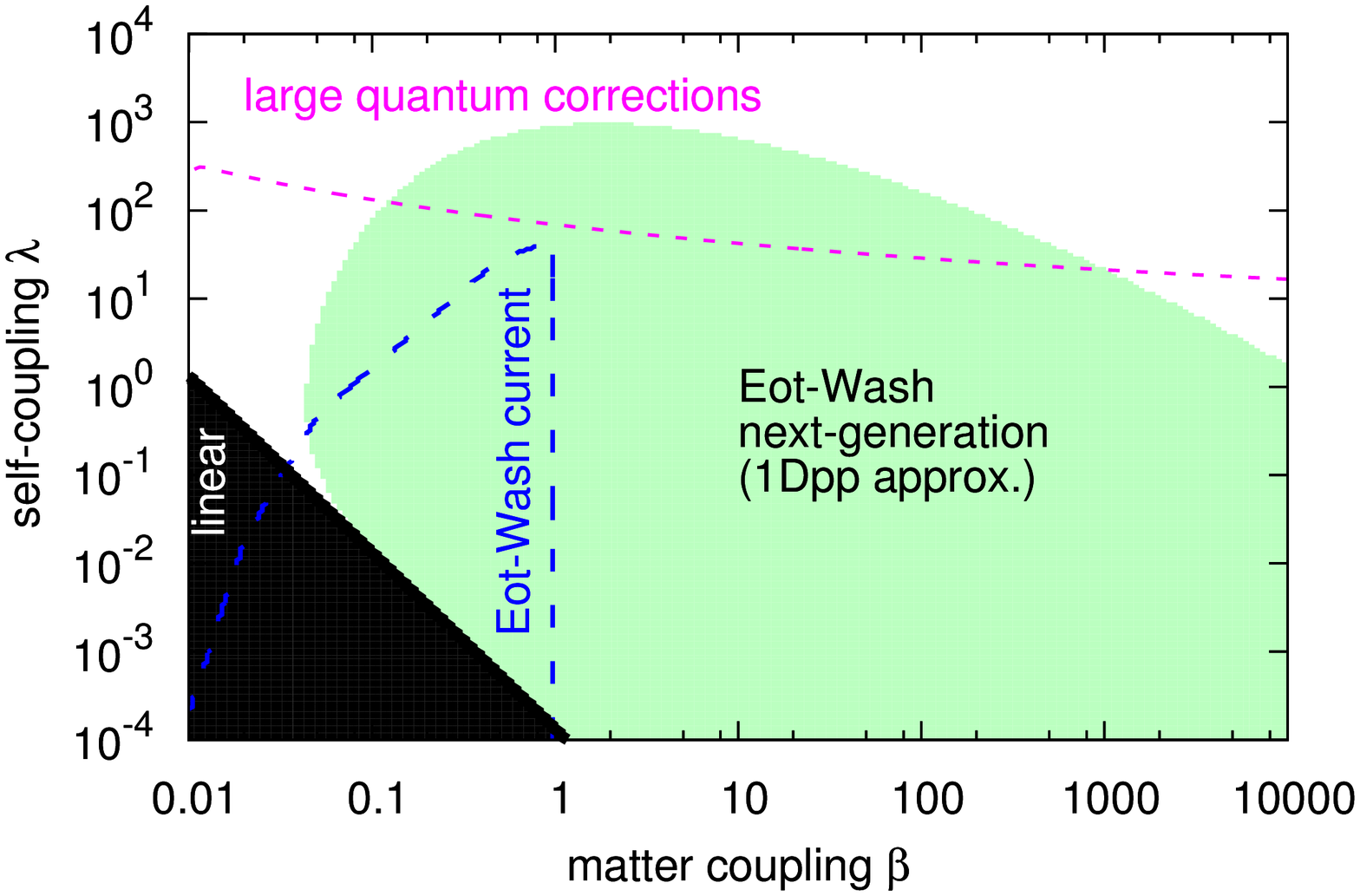}
  \caption{\eotwash~and quantum-stable chameleons.  
    ({\em Left}): Allowed, quantum-stable chameleons lie in the narrow white region between models with large loop corrections and those currently excluded.
    ({\em Right}): The next-generation \eotwash~experiment will exclude all quantum-stable $\phi^4$ chameleons with matter couplings between $0.1$ and $1000$.
    \label{f:quantum_stability_and_eotwash}}
\end{figure}

One-loop quantum corrections $\DVloop(\phi) = \meff^4/(64\pi^2) \log(\meff^2/\mu^2)$ modify the self-interaction $V(\phi)$, altering $\phibulk$, $\meff$, and the predicted fifth force.  In ``quantum-stable'' chameleon models, these corrections remain small inside laboratory fifth force experiments~\cite{Upadhye_Hu_Khoury_2012}.  Neglecting the logarithm in $\DVloop$, the quantum stability condition is $\meff < 0.0073 (\bmat\rho/ 10\text{ g cm}^{-3})$~eV.  Models with large quantum corrections in a laboratory experiment with $\rho = 10$~g/cm$^3$ are shaded in the upper left hand corner of Fig.~\ref{f:quantum_stability_and_eotwash}~({\em Left}).  The shaded region in the lower right approximates the constraints of \eotwash~\cite{Kapner_etal_2007}, leaving a small allowed region near $\bmat \sim 1$.

These allowed models near $\bmat \sim 1$ can be excluded by the next-generation \eotwash~experiment, which will be several times more sensitive to chameleon fifth forces than the current apparatus~\cite{Upadhye_Gubser_Khoury_2006}.  Using a one-dimensional plane-parallel (1Dpp) calculation which approximates the geometry locally as an exactly-solvable one-dimensional configuration, Refs.~\cite{Upadhye_2012a,Upadhye_2012b} estimate fifth forces and constraints.  Figure~\ref{f:quantum_stability_and_eotwash} shows that the next-generation \eotwash~experiment will substantially improve upon current constraints~\cite{Adelberger_etal_2007} and will exclude a large range of quantum-stable chameleon models.

\section{New particles}
\label{sec:new_particles}

Next we consider nonzero $\bgam$.  The equation of motion $\Box\phi = \Veff'(\phi)$ and the modified Maxwell equations $\partial_\mu [ (1 + \bgam \phi / \Mpl)F^{\mu\nu}]=0$ couple the scalar and electromagnetic fields such that, in a background magnetic field $\vec B_0$, a photon may oscillate into a chameleon particle.   The oscillation amplitude is proportional to $\bgam$ and $B_0$. For $\meff^2 \ll 4\pi \omega / L$, where $\omega$ is the photon energy and $L$ is the length of the magnetic field region, the amplitude is also proportional to $L$.

\begin{figure}[tb]
  \includegraphics[angle=270,width=2.8in]{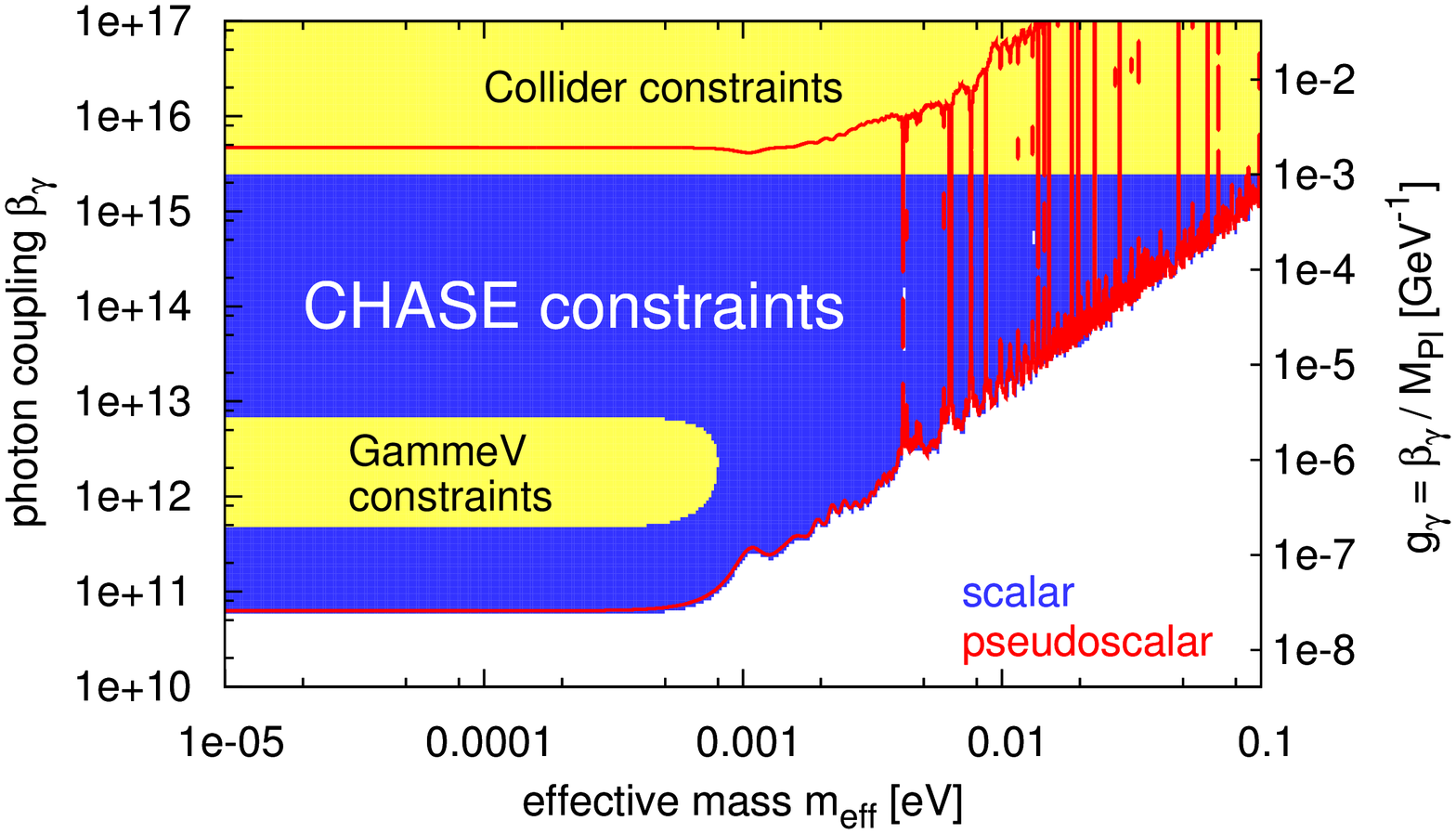}
  \includegraphics[angle=270,width=2.8in]{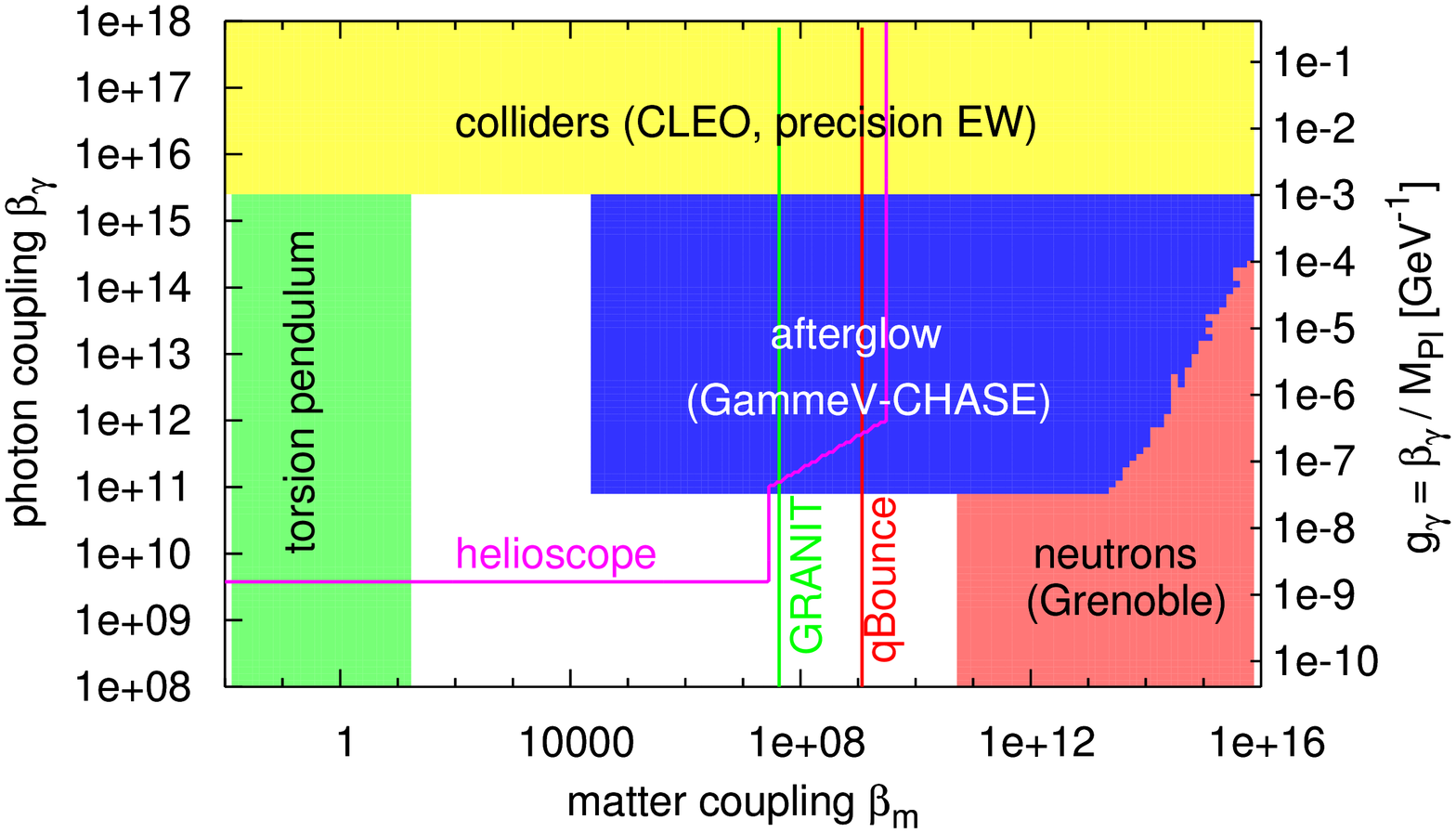}
  \caption{Constraints on photon-coupled chameleon dark energy.
    ({\em Left}): Model-independent CHASE constraints.
    ({\em Right}): Combined constraints for $V(\phi) = M_\Lambda^4(1 + M_\Lambda/\phi)$.  Constraints from colliders~\cite{Brax_etal_2009} and neutrons~\cite{Brax_Pignol_2011} as well as forecasts for helioscopes~\cite{Brax_Lindner_Zioutas_2012,Baker_etal_2012a} are shown.
    \label{f:afterglow_constraints}}
\end{figure}

If the magnetic region is bounded by dense walls inside which $\meff \gg \omega$, then by energy conservation the chameleon particles will be trapped inside the magnetic region. An afterglow experiment streams photons through a dense-walled vacuum chamber enclosing a magnetic field in order to build up a population of trapped chameleon particles.  After the photon source is turned off, these chameleons regenerate photons which may emerge as a detectable afterglow.

A thorough study of the design and analysis of afterglow experiments is given in~\cite{Upadhye_Steffen_Chou_2012}.  Monte Carlo simulations are used to compute the rate at which the trapped chameleon population decreases, and the rate at which detectable afterglow photons are regenerated.  Glass windows inside the magnetic field chamber, used by CHASE to lessen the effects of destructive interference at large $\meff$, are shown to mitigate the adiabatic suppression of photon-chameleon oscillation.  Systematic effects, such as transient glows emitted by vacuum materials in CHASE~\cite{Steffen_etal_2012}, are also considered.  Model-independent CHASE constraints are shown in Fig.~\ref{f:afterglow_constraints}~({\em Left}), while Fig.~\ref{f:afterglow_constraints}~({\em Right}) compares CHASE constraints on inverse power-law chameleon dark energy to the \eotwash~analysis of Sec.~\ref{sec:fifth_forces} as well as constraints from colliders and ultracold neutrons.

\section*{Acknowledgments}

The author thanks O.~K.~Baker, C.~Burrage, A.~Chou, A.~Lindner, J.~Steffen, W.~Wester, and K.~Zioutas for insightful discussions.
The author was supported by the U.S. Department of Energy, Basic Energy Sciences, Office of Science, under contract No.~DE-AC02-06CH11357.

The submitted manuscript has been created by
UChicago Argonne, LLC, Operator of Argonne
National Laboratory (``Argonne''). Argonne, a
U.S. Department of Energy Office of Science laboratory,
is operated under Contract No.~DE-AC02-06CH11357. 
The U.S. Government retains for itself,
and others acting on its behalf, a paid-up
nonexclusive, irrevocable worldwide license in said
article to reproduce, prepare derivative works, distribute
copies to the public, and perform publicly
and display publicly, by or on behalf of the Government.

\bibliographystyle{unsrt}

\bibliography{chameleon}

\end{document}